# Exploring individual differences through network topology

Yuval Samoilov-Katz[1,2]*, Yoram Louzoun[1,2]*, Lev Muchnik[3], Adam Zaidel[1+]

[1]Gonda Multidisciplinary Brain Research Center, Bar Ilan University, Ramat Gan 5290002, Israel

[2]Department of Mathematics, Bar-Ilan University, Ramat Gan 5290002, Israel

[3]School of Business Administration, The Hebrew University of Jerusalem, 91905, Israel.

*These authors contributed equally towards this study

[+]Corresponding author:
Adam Zaidel
Gonda Multidisciplinary Brain Research Center
Bar Ilan University
Ramat Gan, 5290002
Israel
Tel: +972-3-738-4431
Email: adam.zaidel@biu.ac.il

**Keywords:** personality traits; the big five; social networks; network analysis; deep learning

**Financial disclosures:** None to report

**Funding sources:** This research was supported by the Israel Science Foundation (ISF, grant No. 1291/20 to AZ and 1777/17 to LM).

# Abstract

Social animals, including humans, have a broad range of personality traits, which can be used to predict individual behavioral responses and decisions. Current methods to quantify individual personality traits in humans rely on self-report questionnaires, which require time and effort to collect, and rely on active cooperation. However, personality differences naturally manifest in social interactions such as online social networks. Here, we explored this option and found that the topology of an online social network can be used to characterize the personality traits of its members. We analyzed the directed social graph formed by the users of the LiveJournal (LJ) blogging platform. Individual user personality traits, inferred from their self-reported domains of interest (DOIs), were associated with their network measures. Empirical clustering of DOIs by topological similarity exposed two main self-emergent DOI groups that were in alignment with the personality meta-traits *plasticity* and *stability*. *Closeness*, a global topological measure of network centrality, was higher for bloggers associated with *plasticity* (vs. *stability)*. A local network motif (a triad of 3 connected bloggers) also separated the personality meta-traits. Finally, topology-based classification of DOIs (without analyzing the blog content) attained > 70% accuracy (average AUC of the test-set). These results indicate that personality traits can be detected in social network topology. This has serious implications for user privacy. But, if used responsibly, network identification of personality traits could aid in early identification of health-related risks, at the population level.

## Introduction

Each individual (human and animal) has his/her own unique characteristics, formed by a combination of both nature and nurture[1,2]. Availability of detailed individual-level data has demonstrated the value of personalized treatment in numerous domains[3–5], ranging from medicine, marketing, education, consumption of news, digital and retail products. Simultaneously, the discovery of the effectiveness of personalized treatment has led to a surge in the concern for individual privacy and privacy-oriented research. Extensive research has demonstrated over the years that a number of personal characteristics are closely associated with behavior and can be used to explain the response to interventions. However, quantifying these traits remains a labor-prone task with numerous privacy implications.

Many current methods for quantifying personality differences across individuals rely on self-reported measures (such as questionnaires), or they are subject to lab-based conditions (which are often conducted over small samples). Accordingly, collection of such data requires significant effort, and its scale is limited[6–8] thus narrowing their use in practice.

The development of modern computational tools combined with the exponential growth of data in recent years allows for the large-scale analysis of individual differences, without the need for explicit questionnaires[9,10]. Specifically, network analyses using big data, applied in several recent animal experiments, have yielded important insights, inferred directly from behavior. For example, different brain activity patterns in zebrafish have been linked to inter-individual behavioral differences[11], and individual traits in mice have been linked to gene expression in the brain[12]. Also, two recent studies of rock hyrax natural social networks found that specific network properties (centrality and ranking) were related to longevity and mating success[13,14]. This shows that non-questionnaire-based approaches can be used to infer individual differences (including personality differences) which can also be applied to humans. So too, the goal of studying natural human behavior using large data could be further advanced by network analysis techniques.

Despite the abundance and richness of the available data, it rarely contains direct information on personal traits, thus limiting the application of the available theory for practical purposes. Some methods to relate observable user-generated data to personal traits were developed in recent years. For instance, one could apply natural language processing techniques to reveal individual traits with relatively high accuracy[15]. Here we suggest exploring this issue by mining online social networks for associations between their topology and personal attributes.

Social network formation is governed among other factors by the activity of the individuals comprising the network, their interests, demographics, and other attributes[16–18]. One can expect that the topology of online social networks retains at least part of that information and that it can be mined. Several network properties were previously associated with specific personality traits[19–21]. However, these studies have typically relied on questionnaires and were performed in small networks. The need to investigate whether this information can be extracted from large-scale networks without self-reported measures still lacks a good solution.

In online social media platforms, humans interact by posting their opinions and sharing information from other users and from around the web. To optimize information exchange, they develop and maintain social networks thus altering the information they are exposed to and the reach of the content they publish. Finally, to establish a solid online presence, social media platform users publish details about themselves on their profile pages. Together, these data provide a rich source of information about the relation between personality traits, content, and network structure.

In this study, we analyzed user profiles and social network data from LiveJournal (LJ)– a large-scale social network of bloggers[22,23]. To analyze the LJ network, we defined the bloggers as nodes, connected by directed and unweighted edges according to the bloggers' friendships. Blogger's domains of interest (DOIs, a free text keyword or phrase the blogger chooses to describe his/her interests) were associated with personality traits. This was done based on previously described relationships between keywords/free text (here, DOIs) and personality traits[24–27].

We further used this relationship to explore whether individual user's personality traits, as inferred from their DOIs, are evident in LJ network. For that, we leverage LJ structure so that each node (blogger) was characterized by a set of 20 topological features (in a network attribute vector approach[28] - NAV). An empirical clustering of the correlations of DOIs with these 20 NAV features, from an unsupervised perspective, yielded two main DOIs groups. We found that these groups were in alignment with personality meta-traits - *plasticity* and *stability*. *Plasticity* (vs. *stability*) associated bloggers (by DOI proxy) had larger values of *closeness*, a global topological feature of network centrality. This trend was also evident in a local network triad (a specific motif of 3 connected bloggers), that correlated with *closeness*. Finally, we demonstrate that DOIs are related to the network structure, and enable the classification of DOIs. Using a graph-based learning approach based on network topology and label propagation alone (without using the actual blog content), we attained accuracy (by AUC of test-set) exceeding 70% over the first and final snapshots. Analysis of sequential network snapshots (first and final snapshots) helped to further increase the scale of the training set and validated our method.

# Methods

## Network data

The social network analyzed in this study was the LiveJournal (LJ) blogging platform[22,23] which is a free online platform that allows bloggers to communicate and share information in the form of distinct text entries (stories, ideas, etc.). LJ is a large network, which at the time of this study had around 10 million bloggers with over 120 million friendships among them. LJ friendships are not necessarily symmetric as each blogger can select the profiles he/she follows independently.

Blogger profiles contain a section with a list of self-declared domains of interest (DOI). We used these details to characterize each blogger's main personality traits[24–27]. For that, from the list of 100 most popular DOI's, we extracted 94 which were in English.

The structure of LJ is dynamic, and can change over time - i.e., DOIs and/or edges (friendships) can change (created or eliminated) at a particular point in time. For computational reasons, we analyzed the first and final snapshots of LJ (out of the sequence of 19 snapshots collected over the course of about 22 months). We mainly focused on the final snapshot (unless explicitly stated), since it reflects a more evolved version of the LJ network.

## Individual differences and free text

One of the popular frameworks used to characterize human personality uses the Five-Factor Model to produce five key attributes designated as "the big five"[29]. The big five personality traits can be split into two personality meta-traits, *stability* and *plasticity*[30,31]. The *stability* meta-trait reflects an individual's tendency to restrain potentially disruptive behavior. It comprises a combination of three traits from the big five personality dimensions: *conscientiousness* (an individual's degree of organization and persistence), *agreeableness* (an individual's degree of kindness), and *neuroticism* (reversed; the degree to which an individual experiences the world as distressing, threatening, or unsafe). The *plasticity* meta-trait reflects an individual's tendency for exploration. It comprises a combination of the other two traits from the big five personality dimensions: *openness to experience* (an individual's degree of curiosity) and *extraversion* (an individual's degree of assertiveness and dominance).

The lexical hypothesis proposes that an individual's choice of language is indicative of his/her personality traits[24]. Specific keywords have been previously associated with the big five personality traits[26,27]. It was shown that an online profile is reflective of user's real "offline" personality[32,33]. Hence, for this exploratory analysis, these keywords can be used as proxies for individual personality differences, when studying a network. Accordingly, we used these associations to couple DOIs from the network with the personality meta-traits (Table 1; Not enough DOIs are associated with each of the big five personality traits with high enough accuracy to study them individually).

**Table 1: Associations between personality meta-traits, the Big Five traits, and DOI keywords.**

| Meta-trait | Personality trait | DOI |
|---|---|---|
| *Stability* | *Conscientiousness* | T.V., Cats, Movies |
| | *Agreeableness* | Drinking, Laughing, Sex |
| | *Neuroticism* | Sleep, Sleeping, Life |
| *Plasticity* | *Openness to experience* | Love, Poetry, Literature |
| | *Extraversion* | Internet, Drawing, Fantasy, Books, Video Games, Anime, Manga, Reading, Computers |

## Network measures

A network can be projected onto a graph consisting of nodes (also known as vertices, their count represented by V) and edges (count represented by E) that connect pairs of nodes. Nodes were defined by the bloggers, and edges were defined by (directional) friendships between bloggers. When a blogger had marked someone as a friend, an edge was defined ‘out’ from that blogger, ‘in’ to his/her friend. When available, the directionality and weights of an edge (i.e., with a beginning and an end node) carry additional information. Here, we studied directional edges, assuming all friendships are equal, i.e., unweighted edges. Accordingly, we projected the LJ network onto a directed and unweighted graph.

To learn the association between the blogger’s personality traits and the topology of his/her social network, we analyzed each blogger’s social network to compute a list of network measures. This list of features packed in a vector is sometimes designated as the Network Attribute Vector[34] (NAV) and reflects both local and global network topology associated with the specific user. The following features were used for each blogger:

(a) *In-degree, out-degree*: the number of other nodes that are connected to (in) or from (out) the measured node. *In-degree* is proportional to how many times the blogger was chosen as a friend, and the *out-degree* is proportional to how many friends the blogger had chosen to follow.

(b) Four node centrality measures: *Closeness*, *k-core,* and *breadth-first search* (*BFS*) first and second moments.

*Closeness* (formally known as *closeness centrality)* is the average length of the shortest path between the node and all other nodes in the graph[35]. *Closeness* is sometimes associated with the rate at which the information originating at a node can reach the entire network[36].

*K-core* is a measure of a node's degeneracy. This feature reflects the maximal degree ($k$) for which the node remains a member of a $k$-degenerate subgraph (one in which all nodes have $k$ or more edges) and can help ranking nodes by access to other central nodes in the network[37].

*BFS* first and second moments are the average and the standard deviation of the BFS distance distribution (respectively).

(c) *PageRank:* the probability of reaching the node by a random walk with episodic random relocations, when starting at a random location. This is a global measure, that models navigation over the network and was used by Google to rank nodes, by the probability that they are the target of one's search[38,39].

(d) *Motif3*: representing the number of times that the node participated in a unique (specific) subgraph of three connected nodes (triads). Over (or under)-represented triads hint upon the processes guiding network evolution and its functionality. M*otif3-x* represents thirteen possible triad combinations in directed graphs and contributes 13 features to the user's NAV. We defined the specific triad numbers according to a recent study[40] which highlighted their importance over diverse real-life networks, as can be seen in Supplementary Figure 1.

Together these measurements comprise the 20-features-long NAV.

## Network analysis

First, we uniformly sampled the original network over ~$2\times10^6$ nodes (bloggers), taking the largest connected component (Fig. 1a). This reduction (from $1\times10^7$ bloggers to ~$2\times10^6$ bloggers) was done for computational reasons. It was recently shown that this type of random selection of nodes is less biased on larger samples[41].

This resulted in a sample of V = 1,218,319 unique bloggers (with E = 7,225,480 links defined by friendship relations). We then calculated the NAV for each blogger in the sampled network, forming a V (blogger) × 20 (NAV features) matrix. This network features matrix is complemented with a V × 94 matrix in which each row represents one-hot encoded (binary) DOI's of the corresponding blogger (Fig. 1b).

To understand the relationship between the user network properties and his/her domains of interest, we calculated a network topology correlation matrix. Each cell of this 94 x 20 matrix contains the correlation between one DOI and one social network feature (Fig. 1c). Considering that the relationship between many of the pairs of these variables is not linear and many of the network measures are heavy-tailed[42], we used Spearman's rank correlation rather than the more common Pearson[43] (calculated using the SciPy python library[44]).

The importance of topological features varies among different social networks[34]. To analyze and compare the topological profiles of DOIs in this sampled network, the correlation matrix was reduced to a low-dimensional representation. The resulting correlation matrix reflects the extent to which each NAV feature is associated with the DOI of interest. This captures the broader social network aspects of DOIs in a concise form. The main advantage of using this correlation matrix (vs. the raw NAV features[34,45–47]) is that it allows us to compare topological features in a normalized manner. By contrast, the NAVs contain a variety of features that can be discrete or continuous and with or without bounds. And, the distribution of discrete variables in networks (e.g., in- and out-degrees) is often heavy-tailed.

To map the relationship between DOIs and NAV features, DOIs were grouped by topological similarity by optimization of minimum variance between clusters using agglomerative hierarchical clustering of the correlation matrix (Fig. 1d, bottom left). Then, we analyzed the different DOIs using principal component analysis (PCA; Fig 1d, bottom right). For an illustration of the distribution of DOIs over the PCA space, the eight most popular DOIs are represented by the blue labels on the right plot. This analysis was done to identify the major topological features across DOIs, i.e., features that carry most of the DOIs-related information.

## Neural network

To demonstrate that this relationship can be used to decode DOIs, we implemented an artificial neural network (ANN) for the classification of DOIs, based on the raw NAV features. The goal was to train a classifier to predict DOIs by their network properties and by label propagation. We implemented a binary classifier – specifically a feedforward neural network composed of three layers. This was defined similarly to our recent publication[34]. However, we implemented two minor adjustments to prevent overfitting when optimizing the parameters. First, we applied a penalty to prevent the weight matrices from being too large (regularization). We used L2 regularization - with a 0.1 rate regulation (penalty) for the first layer, and 0.01 rate regulation for the second layer. We used a dropout rate of 0.2 for both the input layer and the hidden layer.

## Input to the neural network

Classification of vertices in networks can be done through two main mechanisms: (i) Label propagation, based on the concept that neighboring nodes have similar labels (here, DOIs), and (ii) topological information. Label propagation is useful for node classification[48], especially when combined with network topology[34,46]. Thus, for DOI classification we used a combination of network topology (using all 20 NAV features) and label propagation (4 additional features, as described below in further detail). Both were calculated separately over the first and final snapshots of the network ($1^{st}$ and $19^{th}$ snapshots accordingly).

For label propagation, the sampled network was first divided into separate training and test sets, comprising 80% and 20% of the nodes, respectively. To measure label propagation for the training set, we computed the number of first neighbors each blogger had with and without a given DOI. Specifically, for each DOI and for each node, we counted the number of nodes pointing to it with and without this DOI, and the number of nodes he points to with and without this DOI. This process resulted in four features per DOI, which were added (to topological information) as input to the ANN.

## Data analysis and statistics

For the clustering and classification analyses, we used the Scikit-learn and Keras (with Theano backend) packages[49,50]. To measure the precision of the classifier for each DOI, the area under the curve (AUC) of the receiver operating characteristic (ROC) curve was calculated on the test set. Statistical tests (*t*-tests and $\chi^2$) were computed using JASP[51] (version 0.13.1.0). Given the exploratory nature of this analysis, raw p-values are reported (no correction for multiple comparison). The datasets generated and analyzed in this study and network analysis code are available from the authors upon request.

# Results

This study aimed at exploring whether individual personality differences can be detected in social network topology, without the use of self-reported personality measures like questionnaires. To do so, we projected a sample of LJ bloggers onto a connected (directional) network, as depicts in Figure 1. The network nodes were defined as bloggers, and edges/links were defined as friendships among them (i.e., an edge exists from blogger 'i' to blogger 'j' only if 'i' had marked 'j' as his/her friend). We analyzed the network by the DOIs from profiles of the LJ bloggers, which reflect different personal interests (see Methods). Using earlier studies, the association of the ninety-four most popular English language DOIs with one of the two personality meta-traits[26,27] (stability and plasticity) were defined (Table 1; we studied the 100 most frequent DOI, but 6 were not in English). This resulted in three groups: stability (N = 9), plasticity (N = 12) and non-associated DOIs (N = 73). Based on these DOIs, we investigated how individual differences in personality traits (by DOI proxy) are represented in the network.

## Topology-based clustering of DOIs reveals two main groups

We first analyzed the network topology correlation matrix to find DOI groups (see 'Network analysis' section in Methods for elaboration). Specifically, we looked at the self-emergent grouping of DOIs by their topological similarity using unsupervised hierarchical clustering[28].

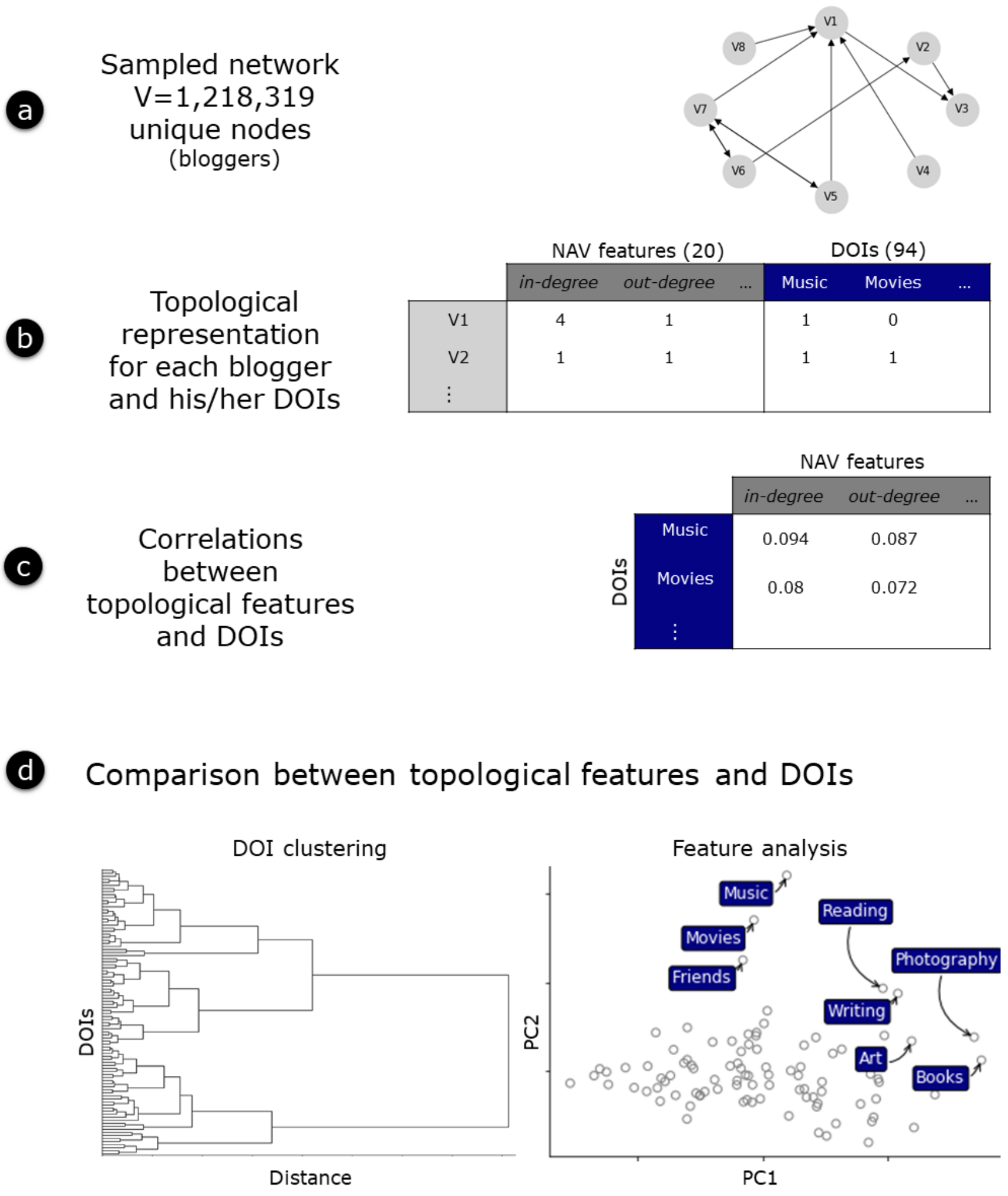

**Figure 1. Network topology analysis stages.** (a) Uniform sample of bloggers from the LiveJournal (LJ) network, taking the largest connected component (resulting in V = 1,218,319 unique bloggers). A sample graph (right) depicts an example subnetwork of LJ network. (b) Each node (blogger) was characterized by (i) a set of 20 topological features (network attribute vector, NAV; marked by dark gray) and (ii) a set of 94 most popular DOIs (domains of interest; reflects the blogger's personal interests; marked by blue) in a binary representation. (c) The correlation of DOIs with NAV features resulted in a matrix of size 94x20, i.e. 20 features for each of the 94 DOIs (by the characteristics of each node from (b)). (d) Analysis of these correlations (as a proxy of network topology) using agglomerative hierarchical clustering (bottom left) and principal component analysis (PCA, bottom right). The blue labels mark the eight most popular DOIs in PCA space.

Figure 2 presents the hierarchical clustering of the 94 most popular DOIs by this correlation matrix. The split of these DOIs reveals two groups – one containing 57 DOIs, and another composed of 37 DOIs. For instance, 'reading' and 'writing' were clustered together in one group, whereas 'football' and 'basketball' in the other.

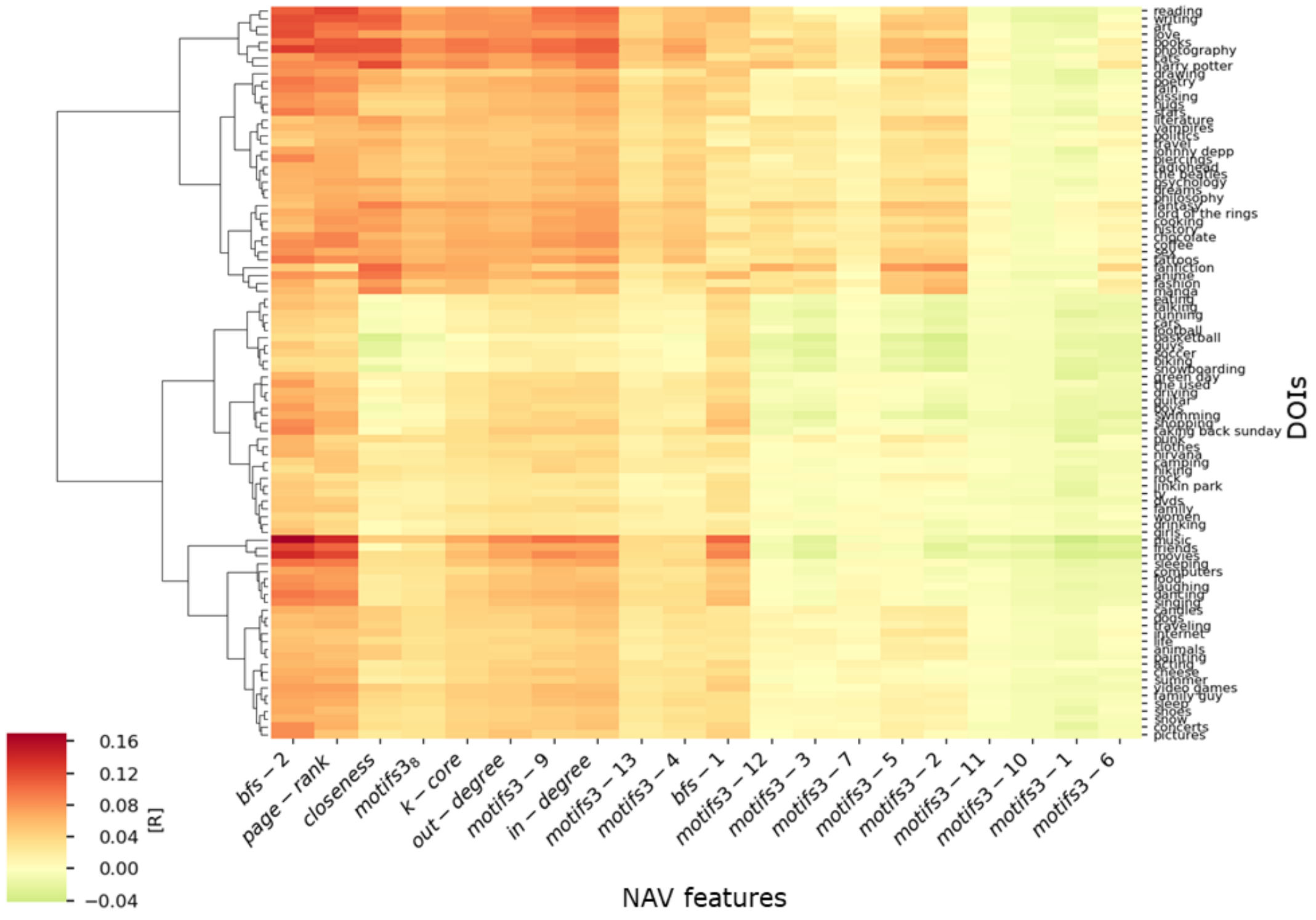

**Figure 2. Hierarchical clustering of the network topology correlation matrix reveals two main DOI clusters.** Correlation (Spearman's ρ) of the 94 most popular DOIs with 20 NAV features.

We then analyzed the extent to which these two main hierarchical clusters match the meta-traits of personality. The split of DOIs by hierarchical clustering and by meta-traits of personality is presented in Table 2. There was a significant relationship between the unsupervised hierarchical clusters (cluster I and II as presented in Fig. 2 by the top and bottom clusters respectively) and meta-traits of personality ($\chi^2$ = 5.743, $p$ = 0.017; Cramer's $V$ = 0.523). These results indicate that topological clustering of DOIs grouped them largely per the meta-traits of personality.

**Table 2: Meta-traits of personality and topological similarities by DOI keywords**

| # | Cluster I | Cluster II | **Total** |
|---|---|---|---|
| *Stability* | 2 | 7 | **9** |
| *Plasticity* | 9 | 3 | **12** |
| **Total** | **11** | **10** | **21** |

## Network topology exposes meta-traits of personality

To find the most influential topological features we projected the DOIs onto a low-dimensional representation of network topology using principal component analysis (PCA). The PCA projection can be seen in Fig. 3, wherein the first PC accounts for 63.0% of the variance. Therefore, we further studied the specific NAV features that contributed most to PC1. To explore these features, we compared the top five topological features (the top five contributors, based on their PC1 weights; Fig. 3b) between *plasticity* vs. *stability* associated DOIs.

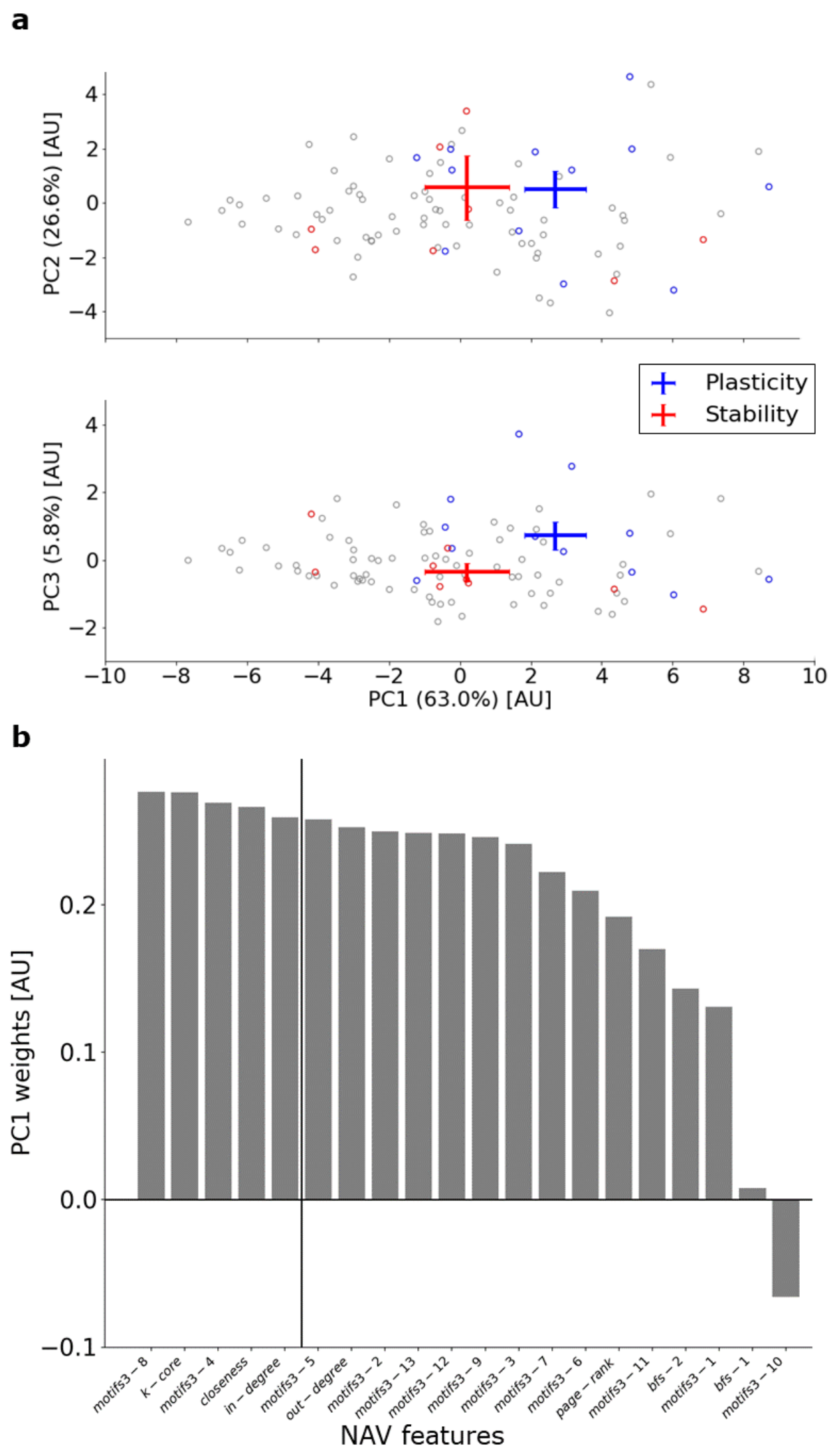

**Figure 3. Personality meta-traits *plasticity* and *stability* over PC space.** Principle component analysis (PCA) was performed on the network topology correlation matrix. (a) Data are presented over PC1×PC2 (top plot) and PC1×PC3 (bottom plot). The percentages in parenthesis present the amount of variance explained by that PC. Each datapoint (circle markers) represents one DOI (per scatter plot). DOIs associated with personality meta-traits *plasticity* and *stability* are marked by blue and red circles, respectively (unassociated DOIs in gray). Plus markers represent the mean values ± SEM for *plasticity* (blue) and *stability* (red). (b) NAV features with highest PC1 weights.

When comparing the top five NAV features of DOIs associated with *plasticity* vs. *stability*, *plasticity* had larger values for *closeness* ($t(19) = 3.130$, $p = 0.006$, uncorrected p-value). *Motif3-8* showed a similar trend ($t(19) = 2.481$, $p = 0.023$, uncorrected p-values). We present the distribution of DOIs for these two network features (*closeness* - the average length of the shortest path between the node and all other nodes; *motif3-8* – three vertices, with two out of three connected by bi-directed edges) in Figure 4a. For illustration, *closeness* and *motif3*_8 values are presented for the example graph in Figure 4b (purple numbers on the nodes, and the specific triad marked by the green ellipse, respectively). Higher values for *plasticity*-related DOIs (blue) compared to *stability*-related DOIs (red) can be seen for both features. *Closeness* and *motif3-8* were themselves highly correlated ($\rho = 0.983$, $p < 0.001$). The association between DOI and stability vs plasticity can thus be seen at the global level (*closeness*), but evident also at the local level (*motif3-8*). The values of *k-core*, *in-degree* and *motif3-4* did not differ between *plasticity* and *stability* ($t(19) = 1.703$, $p = 0.105$, $t(19) = 1.479$, $p = 0.156$, and $t(19) = 1.070$, $p = 0.298$, respectively; uncorrected p-values).

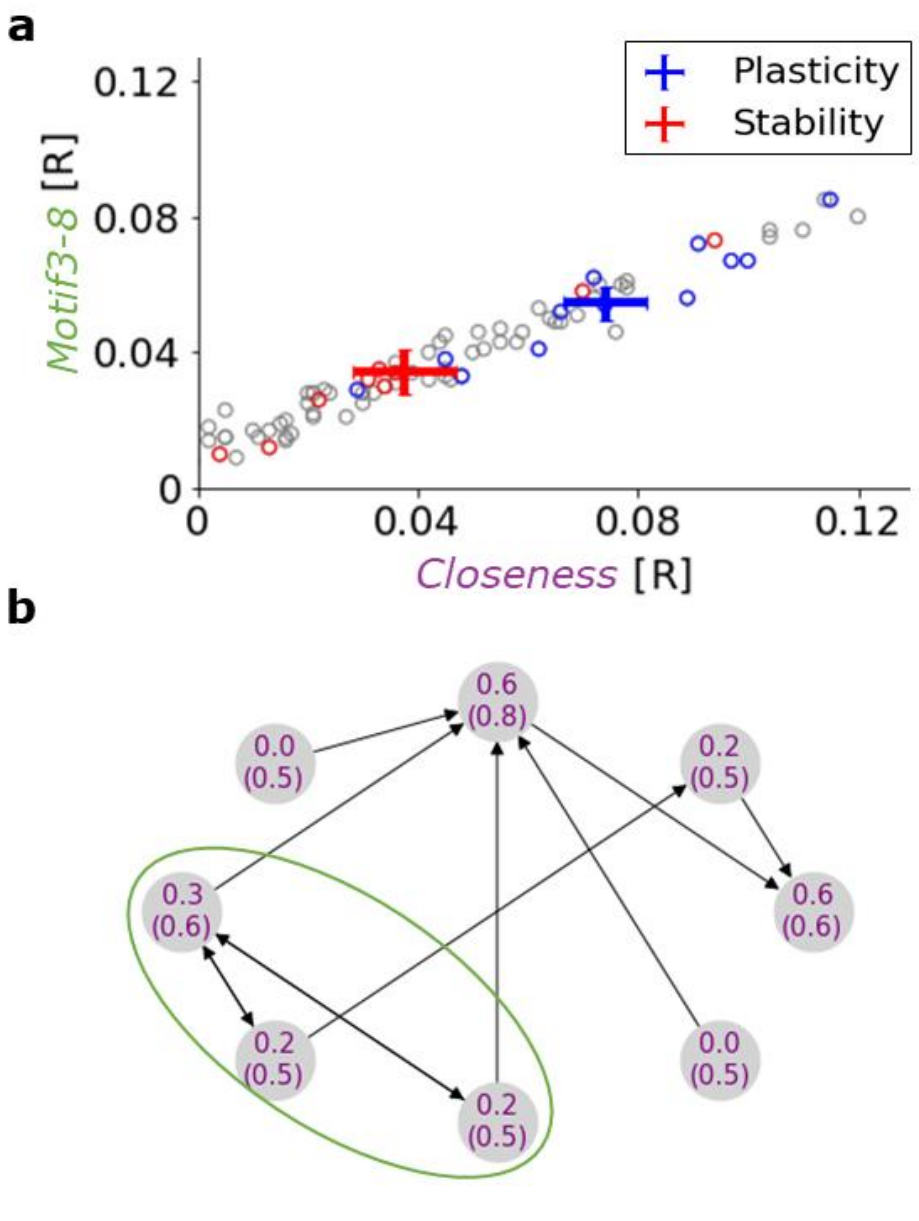

**Figure 4. *Closeness* and the network triad *motif3-8* dissociate personality meta-traits.** (a) Distribution of these two topological features over DOIs (circle markers). DOIs associated with personality meta-traits *plasticity* and *stability* are marked by blue and red circles, respectively (unassociated DOIs in gray). Plus markers represent the mean values ± SEM for *plasticity* (blue) and *stability* (red). Comparing these features between *stability* vs. *plasticity* meta-traits yielded $**p = 0.006$ for *closeness*, and $*p = 0.023$, for *motif3-8* (uncorrected values). (b) *Closeness* scores are presented on the nodes of the example graph from Figure 1a (purple numbers; numbers in parenthesis are *closeness* scores for an undirected graph). An example *motif3-8* subgraph is marked by the green ellipse.

## Topology-based classification of DOIs

Social networks, including LJ, are dynamic and change over time. For example, new DOIs can be added and friendships can change. Current methods for node classification (based on label propagation, i.e. the spread of a specific label/class through the network) use information from their surrounding neighbors[48]. Class propagation is based on the assumption that in analogy with social network homophily[16], neighboring nodes have similar classes. It was recently shown that network topology information is complementary to label propagation features, and that classification accuracy of graph components (nodes/edges) improves when integrating both[34,46]. Also, network topology-based classification is less sensitive to the training set size than propagation-based classification[34]. Accordingly, we classified DOIs using network topology combined with label propagation. This process of classification relied only on those network features and did not take into account the blog's content. Finally, we tested whether classification was robust to network changes, by analyzing the first and final snapshots of the LJ network.

For this, a feed-forward deep learning classifier (see 'Neural network' section in Methods) was applied separately to the first and last snapshots. We chose a training size of 20% (of the sampled network). A total of 24 network measures were used - 20 NAV features and 4 label propagation features (based on first neighbors; see 'Input to neural network' section in Methods).

We classified 43 (of the 94 DOIs; randomly chosen) over the first and last ($19^{th}$) snapshots. On average, the DOI classification accuracy (estimated by the AUC of the test set) was 70.8% ± 2.5% and 72.2% ± 2.8% for the first and last snapshots respectively (mean ± SD; range = [65.6%, 79.3%]). AUC for the eight DOIs from Figure 1d, together with the two DOIs with highest ('manga') and lowest ('internet') AUC are presented in Figure 5a. These data show that topology-based classification of DOIs is possible (compared to a random classification), with above 70% accuracy over these two snapshots ($1^{st}$ and $19^{th}$ snapshots).

Finally, by grouping DOIs according to their associations with personality traits (Table 1), we classified personality using the combination of network topology and label propagation. For *extraversion* (which was associated with more than three DOIs) we randomly selected 3 DOIs from its list ('video games', 'manga', 'fantasy') to be comparable to the others. Practically, we concatenated the DOIs, forming a topological profile of personality by feature dimension. These profiles of personality were then decoded, with an average classification accuracy (AUC of the test set) of 73.4% ± 1.4% and 73.7% ± 2.3% on the 1st and 19th snapshots respectively (Fig 5b; mean ± SD).

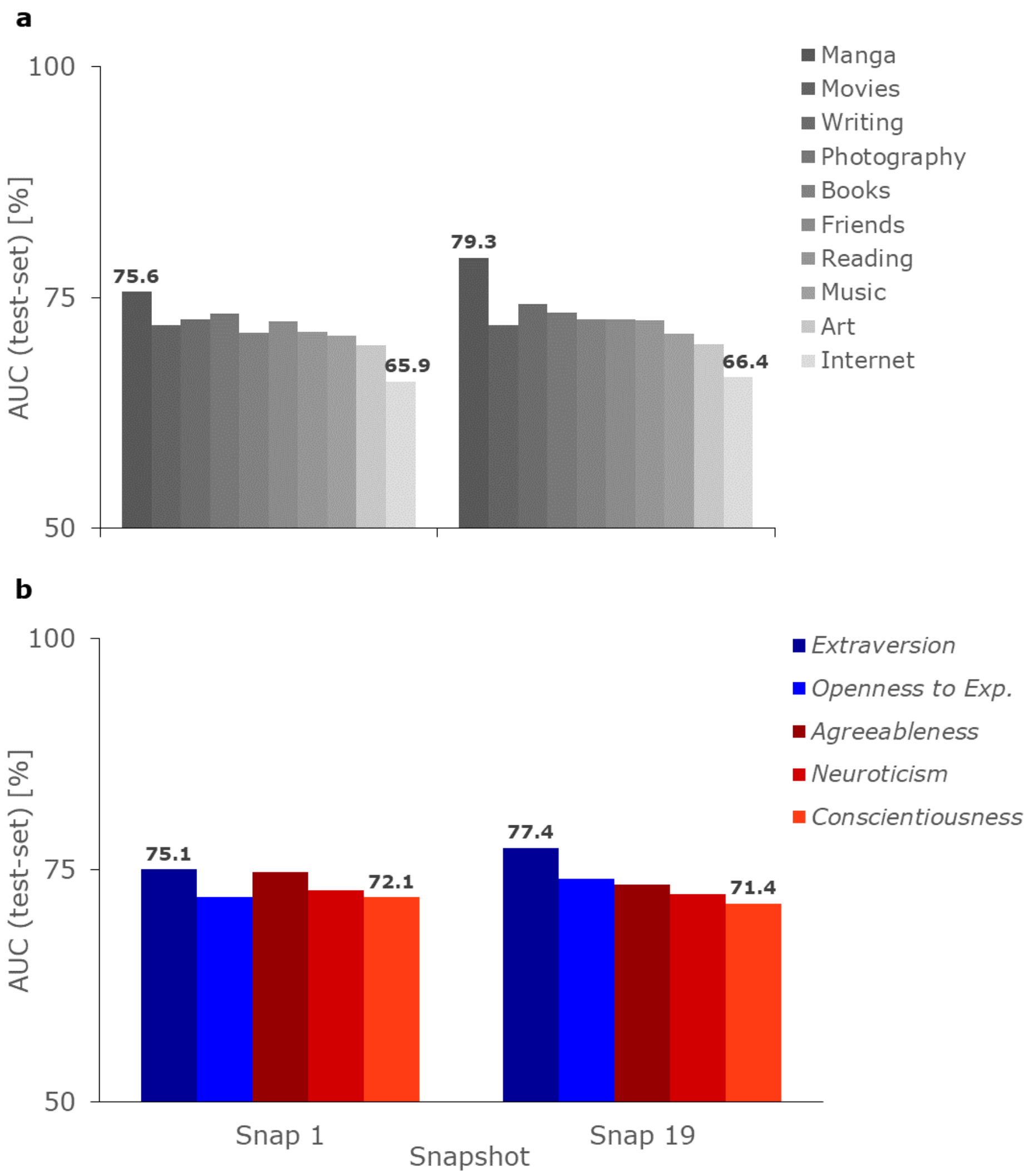

**Figure 5. Topology-based classification of DOIs and personality traits.** Classification of (a) 10 representative DOIs (the eight DOIs from Figure 1d, plus the two DOIs with the lowest and highest classification accuracy) and (b) personality traits using NAV features combined with information propagation features. Accuracy, measured by area under the curve (AUC) of the test set, is presented for two different snapshots of the LJ network (taken about 22 months apart). Each personality dimension was represented by a combination of three associated DOIs.

## Discussion

In this study, we tested the relationship between the structure of LJ, a real-life social network of bloggers, and personality traits. Specifically, we tested whether individual personality differences are empirically evident in the network topology. An important advantage of our approach is that it does not depend on personality questionnaires, which are the primary method to measure personality differences. Rather, we found that personality differences can be exposed in the LJ network through topological features. Hierarchical unsupervised clustering of network topology divided DOIs largely according to their prior associations with personality meta-traits. We found that *plasticity*-related bloggers had larger *closeness* values vs. *stability*-related bloggers. This statistical relationship likely reflects a tendency of individuals with *plasticity*-dominant personality traits to interact more and to take more central positions in the network[52,53]. Hence, we have shown that the huge amount of "raw" behavioral data accumulated on online social networks can be used to identify personality traits.

Local and global topological features were used to measure network topology. Both types of features (local and global) were influential in our analysis (top five). In terms of global features, few recent studies highlighted the importance of centrality measures, beyond simple degree measures, in specific behavioral differences in online social networks (e.g., the judgment of other users by their centrality in the network[34] or identifying influential spreaders[54] by *closeness* and *k-core* respectively). Here, we built on this and demonstrated the importance of *closeness* in identifying personality differences in the LJ network. In terms of local features, *motif3-8* was highly correlated with the global feature *closeness.* A trend for a difference in personality meta-traits was also observed for *motif3-8* (higher values for plasticity). This unique social triad may thus be used as a local proxy of *closeness,* suggesting that more central individuals engage more often in this social triad (m*otif3-8*). This expands the importance of local motifs in real-life social networks[34,40,55]. Note that few recent studies suggest efficient computational alternatives to *closeness,* all are based on local features[54,56,57].

Many personality-related network studies analyze undirected datasets/networks[19–21,58,59]. These ignore directional information. Yet the direction of social interactions (or friendships) is at the basis of real-life communication, i.e., one-way communication is not the same as mutual friendship. This directional information (of communication between humans) is used to learn patterns from human language in state-of-the-art natural language processing (NLP) algorithms (e.g.[60] etc.). In real-life networks, few recent studies also show the importance of using directed measures in both online social[34,46] and biological[40,55] networks.

Inferring personality traits from the structure of small networks may lead to inconsistent results (e.g. *extraversion* and strength of social ties[19,21]) due to *network extraversion bias*[59]. By contrast, large-scale networks offer more consistent, albeit modest correlations, mainly for *plasticity*-related dimensions[31,61]. *Plasticity* reflects an exploration tendency of bloggers, characterized by both behavioral and cognitive aspects[30] (by *extraversion* and *openness to experience* respectively). We found evidence for a combination of those aspects, in more central locations throughout the network. As recently suggested[20,62,63], this might strengthen the existence of *personality homophily*, affecting the underlying structure of online social networks - where individuals with higher levels of *plasticity* (reflecting a higher exploration tendency) are connected to individuals who are more likely to explore ideas with them. This is consistent with efficient knowledge sharing in strategic network locations[52,54].

The differences in personality we found (i.e. higher *closeness* for *plasticity*) might be explained by confounding variables. Demographical information (e.g. age[64] or geographic locations[63,65]) and specific cognitive/social skills (e.g. information seeking preferences[66] and/or relationships quality[67]) should be dissociated in future work. We only analyzed here the correlation between personality traits and network topology (we do not prove a causal relation). Additional evidence for this causal relationship between humans behavior and social online networks was recently shown[17]. We did not study individual items of the big five because there were not enough DOIs associated with each, but our approach provides the basis for future research. Here we identified personality from network topological features without self-reported measures. Future work should specifically label DOIs or text to expand our analysis to the big five, test other (directed) social networks, and introduce additional global and local network measures (not included as part of the NAV features used in this work, such as *betweenness centrality*[68] or *motif4*).

An online social network user does not expect to reveal details about his/her personality, without an explicit approval. These details can potentially be processed to obtain personality profiles for every user, and provide a better understanding of their online behavioral aspects. Such “personality hacking” can be used in negative manners (e.g., mass persuasion by psychological targeting[69] or the Facebook-Cambridge Analytica scandal[70]). Thus, the possibility to accurately extract such details directly from online activity raises privacy concerns. And this should not be neglected by policymakers (or users), who should strive for transparency regarding the utility derived from the collected data (e.g., “personality hacking” warnings, a clear third-party data distribution policy and data compartmentalization could repel violation of user privacy).

From a different point of view, an improved ability to detect a specific personality profile (directly from online behavior in social networks) can be implemented positively across different domains. First, high *extraversion* is associated with an increased risk for accidents[71] (on average they are more prone to take risks) so that accurate detection of this personality profile, can be the target for safety education. Second, a specific combination of the big five dimensions (high *neuroticism* combined with low *extraversion* and low *openness to experience*) was associated with a higher risk of developing Parkinson’s disease[72] (PD) or Alzheimer’s disease[73] (AD). This example is of unique importance, wherein the communality of PD and AD might be used for early identification without self-report measures over online social networks. The clinical evaluation of these neurodegenerative diseases includes elements of self-report measures - part II of UPDRS[74] for PD or MMSE[75] for AD. Therefore, alternatives should be further explored (e.g.[76]) based on real-life behavior and decisions.